# Effect of density on the physical aging of pressure-densified polymethylmethacrylate

R. Casalini and C.M. Roland
Naval Research Laboratory, Chemistry Division, Washington DC 20375-5342

*(July 6, 2017)*

**ABSTRACT**

The rate of physical aging of glassy polymethylmethacrylate (PMMA), followed from the change in the secondary relaxation with aging, is found to be independent of the density, the latter controlled by the pressure during glass formation. Thus, the aging behavior of the secondary relaxation is the same whether the glass is more compacted or less dense than the corresponding equilibrium liquid. This equivalence in aging of glasses formed under different pressures indicates that local packing is the dominant variable governing the glassy dynamics. One consequence is that pressure densification yields a reduction in the glass transition temperature. The fact that pressure densification yields different glass structures is at odds with a model for non-associated materials having dynamic properties exhibited by PMMA, such as density scaling of the relaxation time and isochronal superposition of the relaxation dispersion.

**INTRODUCTION**

The properties of liquids cannot be completely characterized without measurements of their dependence on both pressure and temperature, and in particular the behavior under pressure has provided many insights into the phenomena associated with glass formation [1]. A prominent example is the correlation of various properties with the time scale of molecular motions, as observed under isochronal conditions. In these studies, the primary relaxation time, $\tau_\alpha$, is maintained constant through simultaneous control of pressure and temperature [2], with consequent invariance for many liquids of the dynamic correlation length [3,4], the shape of the relaxation dispersion ("isochronal superpositioning") [5,6], the dynamic crossover [7,8] and for a few cases, the melting line [9]. An interpretation of isochronal invariance of properties comes from molecular dynamics (MD) simulations, which have shown that for a certain class of materials, the behavior, including $\tau$, is governed by the existence of isomorphs [10,11,12]. Isomorphs are state points for which various properties are constant in reduced units. In MD simulations isomorphs are identified from correlation between equilibrium fluctuations of the virial pressure and the potential energy [13]. The usual experimental manifestations of isomorphic state points are isochronal superpositioning [5,6], expressed as invariance of the Kohlrausch stretch exponent at constant relaxation time

$$\beta = f(\tau) \tag{1}$$

in which *f* is a function, and the density scaling relation [1 and refs. therein]

$$\tau = g(TV^{\gamma}) \qquad (2)$$

in which *V* is the specific volume, $\gamma$ a material constant, and *g* a function. All non-associated liquids and polymers tested to date conform to eqs.(1) and (2) [1,14].

A property predicted for materials having isomorphs that has not received experimental attention concerns the behavior in the glassy state: For a jump of the equilibrium liquid to an out-of-equilibrium glass, the effective (fictive) temperature depends only on the final density [15]. This means that the density and hence the ensuing physical aging will be independent of the pressure during vitrification if the glassy state is reached via isomorphic pathways [16]. Since the glass transition temperature by definition is associated with fixed $\tau_\alpha$, it is an isomorphic state point. It follows that the density and physical aging of isomorphic liquids should not depend on the vitrification pressure [16]. The process of forming a glass by application of pressure to the equilibrium liquid is referred to as "pressure densification" [17]; thus, the prediction is that isomorphic liquids cannot be pressure densified.

In this paper we describe measurements on a low molecular weight polymethylmethacrylate (PMMA), a polymer that exhibits isochronal superpositioning (eq.(1)) and conforms to density scaling (eq.(2) with $\gamma$=1.9) [18]. Since these are properties of isomorphic materials, the expectation is that the structure and behavior of glassy PMMA should be independent of the pressure during glass formation; that is, it cannot be pressure densified. We find this not to be the case – the density of the glass is an increasing function of the pressure applied to the liquid while forming the glass. Depending on the magnitude of the vitrification pressure, we obtain (after release of the pressure) glassy PMMA that is either less dense or more dense than the corresponding liquid. This means that physical aging involves either negative or positive changes in mean volume as the glass evolves to equilibrium; however, the dielectric strength and relaxation time of the secondary dynamics decrease during aging, irrespective of the sign of the volume change. Evidently, the properties of the secondary relaxation do not depend on the average density.

**EXPERIMENTAL**

The oligomeric PMMA ($M_w$=1,970 D; polydispersity=1.15) was purchased from Polymer Standards Service and used as received. The dielectric permittivity was measured with a Novocontrol Alpha Analyzer. The sample cell consisted of two parallel plates with a 55 mm Teflon spacer (geometric capacitance = 30.4 pF), encapsulated in a flexible barrier to isolate it from the pressure transmitting fluid (silicon oil). The high pressure apparatus for dielectric measurements was: (i) a high pressure vessel from Harwood Inc.

containing the dielectric cell surrounded by the pressurizing fluid; (ii) an environmental chamber (Tenney Inc.) for temperature control temperature; and (iii) a hydraulic system to generate the pressure. The last consisted of two pumps (Superpressure and Enerpac from Newport Scientific), in combination with an intensifier (Harwood Eng.), which enabled pressures up to 1.4 GPa. The pressure was measured with a transducer (Sensotec) and a pressure gauge (Heise).

Pressure-volume-temperature (PVT) measurements were carried out using a Gnomix apparatus, on a ~1 $cm^3$ cylindrical sample formed under vacuum.

**RESULTS**

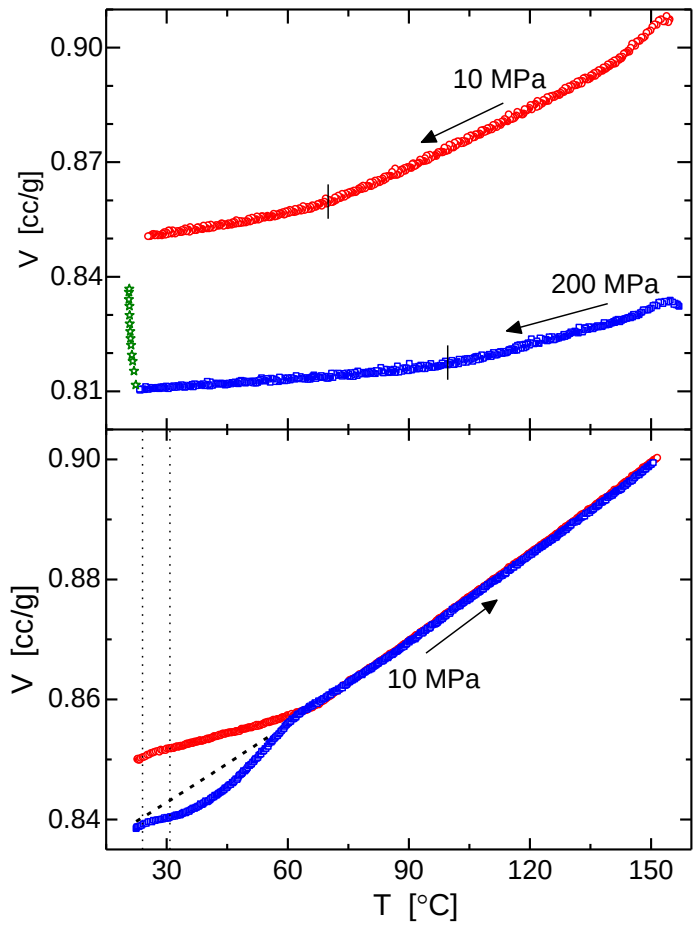

Figure 1. (upper panel) Specific volume during cooling at low and high pressures; vertical data (stars) measured after reduction of pressure from 200 to 10 MPa. Vertical tic marks denote the glass transition temperature. (lower) Subsequent heating curves at low pressure. Dashed line is the extrapolation of the specific volume for the equilibrium liquid. Vertical dotted lines signify the temperatures at which the physical aging was carried out.

The pressure densification method consists of applying pressure to the equilibrium liquid, followed by cooling through the glass transition temperature. The pressure on the glass is then released, so that the temperature and pressure of the material is the same as for the material cooled through $T_g$ at low pressure. Invariably the former is found to have higher density [17], with a metric for the pressure densification defined from the relative volume change [16]

$$\delta = \frac{V_N(P_0) - V_D(P_0)}{V_N(P_0) - V_D(P_1)} \tag{3}$$

in which $V_N$ and $V_D$ are the specific volumes for respective vitrification at low ($P_0$) and high ($P_1$) pressure. Representative results for the PMMA are shown in Figure 1. Note that after release of the pressure, the glass prepared at 200 MPa has a specific volume that is equal to or less than the value for the extrapolated liquid, depending on temperature. This is contrary to the glass prepared at low pressure, which is less dense. Pressure densification also reduces $T_g$, consistent with a more disordered structure.

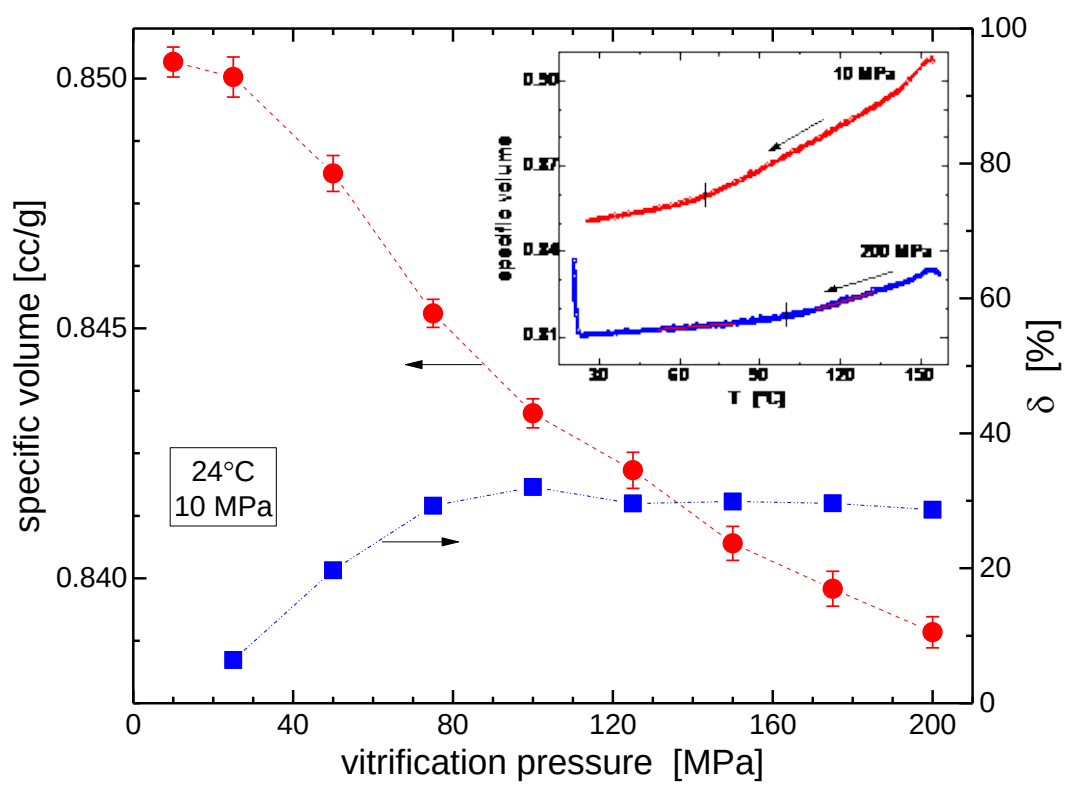

Figure 2. Specific volume (circles) and degree of pressure densification (squares) for PMMA measured at the indicated temperature and pressure as a function of the pressure during cooling from the liquid state.

As seen in Figure 2, the density is an increasing function of the pressure during glass formation. At the highest vitrification pressure, the glass is 5% denser than when formed at the lowest *P*. After release of the pressure, the volume increases but the material remains significantly denser than the ordinary glass; that is, i.e. $\delta > 0$ (Fig 1). Over the range of vitrification pressures used (25 – 200 MPa), $\delta$ varies from ~6% to 29%, which falls in the range of literature results for other polymers [19,20,21]. That PMMA can be pressure densified is at odds with its conformance to isochronal superpositioning (eq.(1)) and density scaling (eq.(2)) [18].

The practical motivation for pressure densification is the expectation that properties can be obtained that will differ from those of an ordinary glass. For example, it has been shown that the mechanical modulus [22], yield strength [23], as well as the structure seen in small angle X-ray scattering [24], are affected by the pressure during glass formation. In this work the property of interest is the structural relaxation time, which generally is too long below $T_g$ to be measured directly. However, we have shown that the changes in the Johari-Goldstein (JG) secondary relaxation are governed by $\tau_\alpha$ [25]. Thus, as aging proceeds, the JG relaxation time, $\tau_{JG}$, increases, with a concomitant reduction in dielectric strength, $\Delta\varepsilon_{JG}$. The assumption of the analysis is that physical aging governs these changes, which occur on a time scale corresponding to $\tau_\alpha$.

Figure 3 shows fits of stretched exponential decay functions

$$\tau_{JG}(t) = \tau_{eq}^{\infty} - A\left(\exp\left[t/\tau_\alpha\right]^{\beta}\right) \tag{4}$$

and

$$\Delta\varepsilon_{JG}(t) = \Delta\varepsilon_{eq}^{\infty}\left(\exp\left[t/\tau_{\alpha}\right]^{\beta}\right) \tag{5}$$

to the respective relaxation time and dielectric strength of the JG relaxation. Here $\tau_{eq}^{\infty}$, $\Delta\varepsilon_{eq}^{\infty}$, , and *A* are constants. For the stretch exponent we take the value measured at $T_g$ (i.e., for the equilibrium liquid), $\beta$ =0.38. As an isolated variable, the densification during physical aging should increase the relaxation time (greater congestion) and the dielectric strength (more dipoles per unit volume). The opposite results reflect the influence that other factors, in particular structure and entropy, exert on the JG dynamics [26,27].

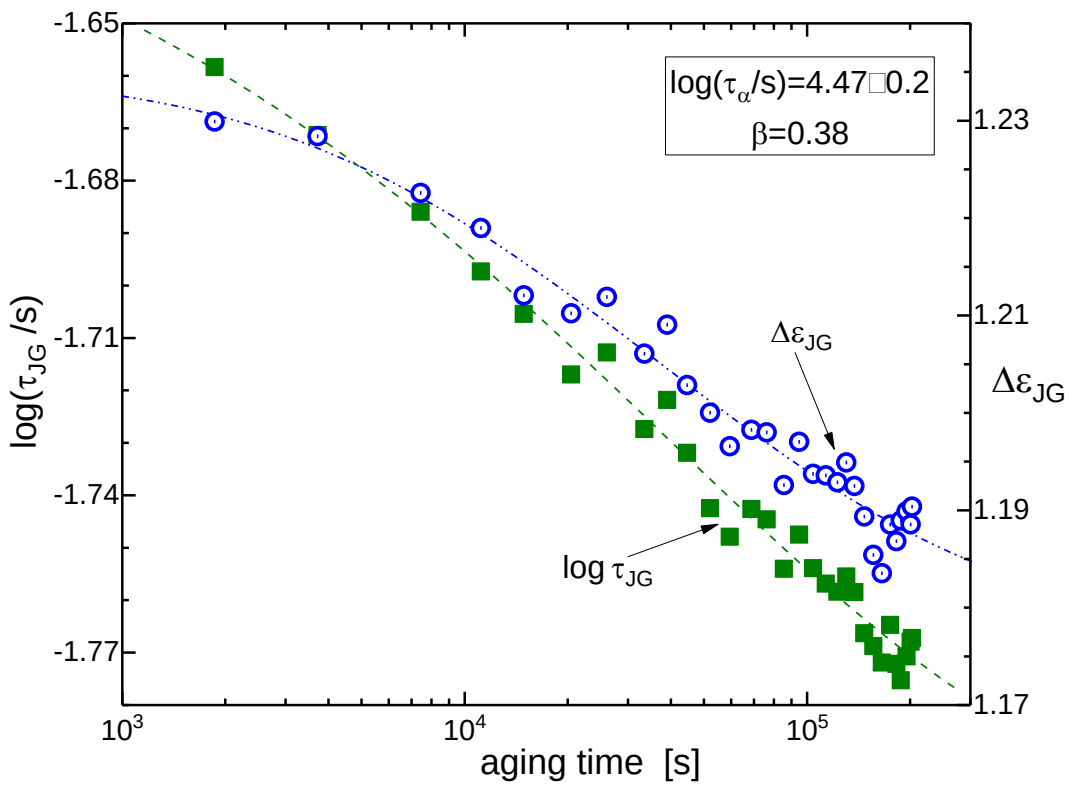

Figure 3. Change in the JG relaxation time and dielectric strength during physical aging of glassy PMMA formed at 188 MPa. The curves are fits to eqs. (2) and (1), respectively, using the Kohlrausch exponent measured at $T_g$ and yielding the indicated value of the structural relaxation time.

From the fits we obtain the α relaxation times shown in Figure 3. This Arrhenius plot includes data for the equilibrium liquid [28]. These $\tau_\alpha$ can be described using an equation due to Hodge [29]

$$\tau_{\alpha}(T) = \tau_{\infty}\exp\left[\frac{B}{T(1-T_0/T_{\mathrm{f}}]}\right] \tag{6}$$

In applying eq. (6), the fictive temperature $T_f$ is set to unity above $T_g$, and then becomes an adjustable parameter for fitting relaxation times in the glass. The parameters obtained for PMMA were log($\tau_\infty$ /s)= 10.61±0.7, *B*=1500±200 K, $T_0$ = 291±4 K, and $T_f$ = 335 K. The temperature dependence show the usual strongly non-Arrhenius above $T_g$ and a weaker, Arrhenius T-dependence in the glassy state.

Since the structure and properties of pressure densified glass are different [22,24], the aging behavior is also expected to differ from that of the conventional glass. However, as shown in Fig. 4, the $\tau_\alpha$ for glassy PMMA, describing the time scale of the physical aging, are essentially independent of the pressure during glass formation. The expectation is that glass-forming materials associated with isomorphs will exhibit simple aging behavior; that is, the aging will be independent of thermodynamic pathway [10,15]. And indeed, we find that the physical aging of the PMMA is independent of its density. The inconsistency is the capacity of PMMA to be pressure densified at all.

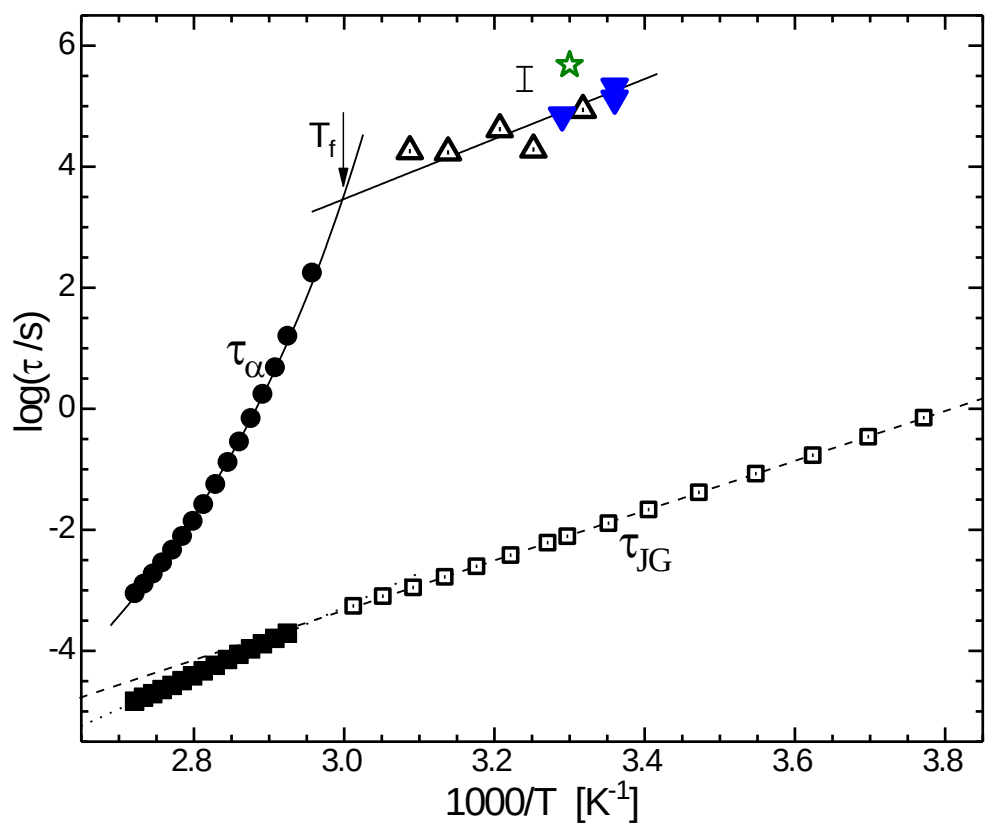

Figure 4. Arrhenius plot of the primary and secondary relaxation times. $\tau_\alpha$ below the fictive temperature (=61.5°C) were obtained from the change of the JG process for PMMA vitrified at 0.1 MPa (triangles) and at higher pressures (185 and 245 MPa; inverted triangles), and from the volume change during aging at ambient pressure (star). The aging was carried out at 24.5°C for glass formation at 185MPa and at 24.5°C and 31.0°C for PMMA vitrified at 245MPa. The curves through $\tau_\alpha$ are the fit of eq.6

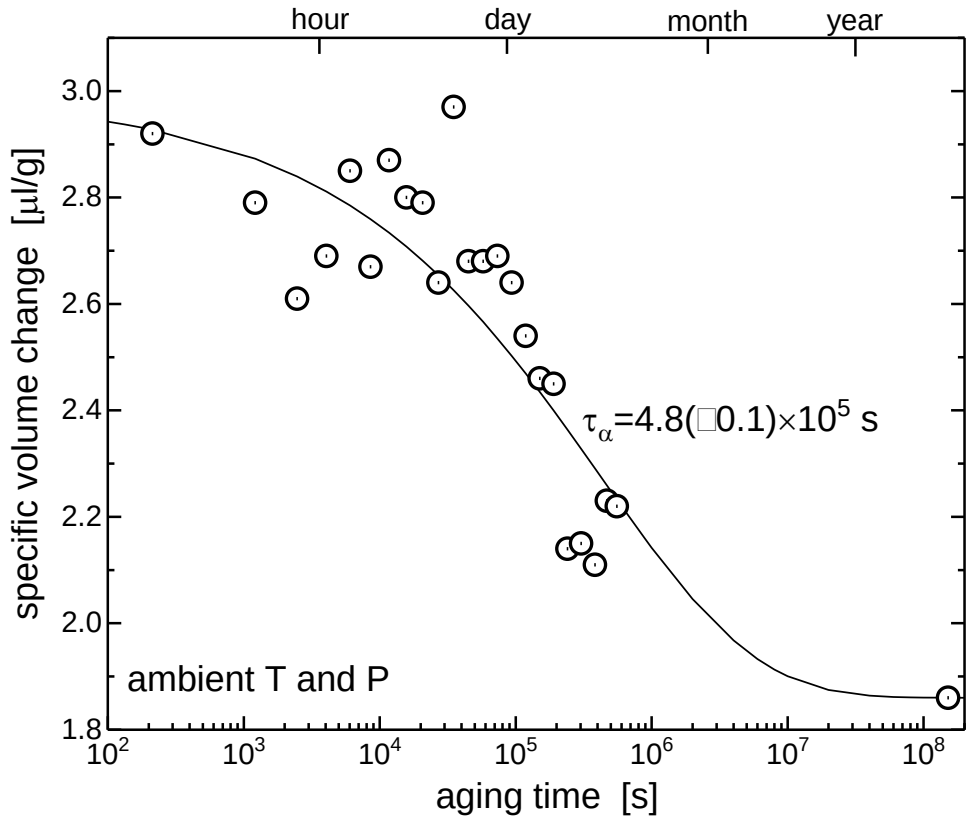

Figure 5. Change in specific volume during physical aging of PMMA vitrified by cooling at 10 MPa.

To verify the $\tau_\alpha$ extracted from the changes in JG properties during aging, the structural relaxation time was measured directly from the change in volume as PMMA evolves toward equilibrium (Figure 5). This experiment included a measurement on a sample aged 4.8 years at ambient conditions. Fitting the data to eqs. (4) or (5) written in terms of *V*, we obtain the value included in Fig. 3. Within the experimental uncertainty, this $\tau_\alpha$ is consistent with the $\alpha$ relaxation times deduced from the change in JG properties.

**CONCLUSIONS**

The pressure during glass formation was adjusted herein so that after its release, the PMMA had a density that was higher, lower, or equivalent to that of the extrapolated equilibrium value. Nevertheless, neither the structural relaxation rate of the glassy PMMA nor the change of JG properties during physical aging were governed by the average density. This result is consistent with pressure densification studies in which glass was prepared having a density equal to that of the equilibrium liquid, but exhibited different distributions of local volume [30]. The average density does not govern the properties of the glass, but rather the local structure and barriers for thermal fluctuations of density are the main control parameters. A more disordered local structure leads to a lower glass transition temperature in the pressure densified polymer. We have previously shown that an asymmetric double well potential model can qualitatively reproduce physical aging behavior, with the degree of asymmetry inversely related to the fictive temperature describing the non-equilibrium structure of the glass [31].

PMMA exhibits properties of a material having isomorphs in its phase diagram; to wit, isochronal superpositioning, density scaling, and as found herein, physical aging kinetics independent of both the conditions during glass formation and the subsequent density. Nevertheless, the density of glassy PMMA is a function of the pressure during vitrification and, since pressure densification follows an isomorphic pathway [16], this pressure dependence is unexpected. Thus, the capacity for pressure densification of PMMA indicates that the polymer has some, but not all, of the properties predicted for materials associated with isomorphs [12]. This apparent contradiction underscores the need for a better understanding of the connection between the properties of real materials and of those found in MD simulations, even for relatively "simple" systems. Particularly for polymers, the properties along isomorphic pathways seems to be related to the flexibility of the polymer backbone [[32]].

**AUTHOR INFORMATION**

*E-mail: mike.roland@nrl.navy.mil (C.M.R.)
E-mail: riccardo.casalini@nrl.navy.mil (R.C.)

**ACKNOWLEDGEMENTS**

Enlightening discussions with D. Fragiadakis are gratefully acknowledged. This work was supported by the Office of Naval Research.

**REFERENCES**

[1] C.M. Roland, S. Hensel-Bielowka, M. Paluch, and R. Casalini, Rep. Prog. Phys. **68**, 1405 (2005).

[2] C.M. Roland, Soft Matter **4**, 2316 (2008).

[3] R. Casalini, D. Fragiadakis, and C.M. Roland, J. Chem. Phys. **142**, 064504 (2015).

[4] D. Fragiadakis, R. Casalini, and C.M. Roland, J. Phys. Chem. B **113**, 13134 (2009).

[5] C.M. Roland, R. Casalini, and M. Paluch, Chem. Phys. Lett. **367**, 259 (2003).

[6] K.L. Ngai, K.L., R. Casalini, S. Capaccioli, M. Paluch, and C.M. Roland, J. Phys. Chem. B **109**, 17356 (2005).

[7] R. Casalini, M. Paluch, and C.M. Roland, J. Chem. Phys. **118**, 5701 (2003).

[8] R. Casalini and C.M. Roland, Phys. Rev. Lett. **92**, 245702 (2004).

[9] D. Fragiadakis and C.M. Roland, Phys. Rev. E **83**, 031504 (2011).

[10] N. Gnan, T.B. Schrøder, U.R. Pedersen, N.P. Bailey, and J.C. Dyre, J. Chem. Phys. **131**, 234504 (2008).

[11] U.R. Pedersen, N. Gnan, N.P. Bailey, T.B. Schrøder, and J.C. Dyre, J. Non-Cryst. Sol. **357**, 320 (2011).

[12] J.C. Dyre, J. Phys. Chem. B **118**, 10007(2014).

[13] U.R. Pedersen, N.P. Bailey, T.B. Schrøder, and J.C. Dyre, Phys. Rev. Lett. **100**, 015701 (2008).

[14] C.M. Roland, Viscoelastic Behavior of Rubbery Materials, Oxford Univ. Press (2011).

[15] N. Gnan, C. T.B. Schrøder, and J.C. Dyre, Phys. Rev. Lett. **104**, 125902 (2010).

[16] D. Fragiadakis and C.M. Roland, submitted; arXiv 1706.05994 (2017).

[17] J.M. Hutchinson in The Physics of Glassy Polymers (R.N. Haward, ed.), Springer (1997).

[18] R. Casalini, C.M. Roland, and S. Capaccioli, J. Chem. Phys. **126**, 184903 (2007).

[19] M. Schmidt and F.H.J. Maurer, Macromolecules **33**, 3879 (2000).

[20] R.E. Wetton and H.G. Moneypenny, Brit. Polym. J. **7**, 51 (1975).

[21] A. Weitz and B. Wunderlich, J. Polym. Sci. Polym. Phys. Ed. **12**, 2473 (1974).

[22] I.V. Danilov, E.L. Gromnitskaya, and V.V. Brazhkin, J. Phys. Chem. B **120**, 7593 (2016).

[23] J.B. Yourtree and S.L. Cooper, J. Appl. Polym. Sci. **18**, 897 (1974).

[24] H.-H. Song and R.-J. Roe, Macromolecules **20**, 2723 (1987).

[25] R. Casalini and C.M. Roland, Phys. Rev. Lett. **2009**, *102*, 035701.

[26] G.P. Johari, J. Non-Cryst. Solids **307**, 317 (2002).

[27] D. Fragiadakis and C.M. Roland, Macromolecules **50**, 4039 (2017).

[28] R. Casalini and C.M. Roland, J. Non-Cryst. Solids **357**, 282 (2011).

[29] I.M. Hodge, Macromolecules **20**, 2897 (1987).

[30] R.E. Robertson, R. Simha, and J.G. Curro, Macromolecules **18**, 2239 (1985).

[31] R. Casalini and C.M. Roland, J. Chem. Phys. **131**, 114501 (2009).

[32] A.A. Veldhorst, J.C. Dyre, and T.B. Schrøder, J. Chem. Phys. **143**, 194503 (2015).